\documentclass[prl,letterpaper,nofootinbib,floatfix,superscriptaddress,preprintnumbers,twocolumn]{revtex4}

\usepackage{color}

\usepackage{graphicx}
\graphicspath{{figures/}}

\usepackage{verbatim}
\usepackage{amsmath}
\usepackage{amssymb}
\usepackage{subfigure}
\usepackage{url}

\usepackage{multirow}
\usepackage{threeparttable}
\usepackage{paralist}

\usepackage{color}
\definecolor{darkgreen}{rgb}{0,0.5,0}

\newcommand{\eV}{\text{~eV}}

\newcommand{\TeV}{\text{~TeV}}

\newcommand{\be}{\begin{equation}}
\newcommand{\ee}{\end{equation}}

\newcommand{\abs}[1]{\lvert #1 \rvert}

\newcommand{\eps}{\epsilon}

\newcommand{\cO}{\ensuremath{\mathcal{O}}}

\usepackage{soul}

\begin{document}

\title{Exponential Hierarchies from Anderson Localization in Theory Space}
 
\author{Nathaniel Craig} \email{ncraig@ucsb.edu}
\affiliation{Department of Physics, University of California, Santa Barbara, CA 93106}
  
\author{Dave Sutherland} \email{dwsuth@ucsb.edu}
\affiliation{Department of Physics, University of California, Santa Barbara, CA 93106}

\date{\today}

\begin{abstract}
We present a new mechanism for generating exponential hierarchies in four-dimensional field theories inspired by Anderson localization in one dimension, exploiting an analogy between the localization of electron energy eigenstates along a one-dimensional disordered wire and the localization of mass eigenstates along a local ``theory space'' with random mass parameters. Mass eigenstates are localized even at arbitrarily weak disorder, with exponentially suppressed couplings to sites in the theory space. The mechanism is quite general and may be used to exponentially localize fields of any spin. We apply the localization mechanism to two hierarchies in Standard Model parameters --- the smallness of neutrino masses and the ordering of quark masses --- and comment on possible relevance to the electroweak hierarchy problem. This raises the compelling possibility that some of the large hierarchies observed in and beyond the Standard Model may result from disorder, rather than order. 
\end{abstract}

\maketitle

\section{Introduction}
It is a truth universally acknowledged, that a quantum field theory in possession of large hierarchies among its parameters must be in want of a deeper underlying explanation \cite{Dirac:1937ti, Dirac:1938mt}. In the context of the Standard Model, considerable effort has been correspondingly devoted to assigning dynamical explanations to observed hierarchies in couplings and scales, such as those observed among fermion masses or in the Higgs sector. Even when hierarchies are rendered technically natural by symmetries, as is the case for Standard Model flavor hierarchies, a truly natural explanation still requires an underlying explanation for the symmetry and the associated dynamical symmetry breaking mechanism. 
 
There are several known mechanisms for generating exponentially suppressed effective couplings or scales in theories without small fundamental parameters. Most notable among these is dimensional transmutation and its higher-dimensional geometric counterparts, which explain the smallness of the proton mass relative to the Planck scale and form the basis for theories of the electroweak scale \cite{Weinberg:1975gm, Susskind:1978ms} and models of flavor in warped extra dimensions \cite{Grossman:1999ra, Gherghetta:2000qt}, among other things. Four-dimensional {\it theory spaces} \cite{ArkaniHamed:2001nc} -- also called {\it quivers} \cite{Douglas:1996sw}, or {\it mooses} \cite{Georgi:1985hf} -- provide another fruitful context for generating large hierarchies. Such theory spaces consist of a series of sites denoting gauge or global symmetry groups, connected by links representing fields transforming under adjacent groups. Examples of hierarchies arising in theory spaces include four-dimensional analogues of volume suppression in flat extra dimensions \cite{ArkaniHamed:2001ca, Hill:2000mu} and wavefunction overlap or coupling dilution in warped extra dimensions \cite{Randall:2002qr, Giudice:2016yja}, as well as more exotic phenomena such as clockwork \cite{Choi:2014rja, Choi:2015fiu, Kaplan:2015fuy, Giudice:2016yja}. But it is notable that generating exponential hierarchies of couplings in theory space typically involves either elaborate order or exponential hierarchies in the fundamental parameters. Given the absence of experimental evidence thus far for elaborate order underlying the structure of the Standard Model, it is natural to wonder if hierarchies of the Standard Model might instead have a more disordered origin.

In this work we present a new mechanism for generating exponential hierarchies in four-dimensional field theories closely related to the phenomenon of Anderson localization \cite{PhysRev.109.1492} in one-dimensional systems. In particular, we exploit an analogy between the localization of electron energy eigenstates along a one-dimensional wire in the presence of disorder, and the localization of mass eigenstates along a local theory space with random mass parameters on each site.\footnote{Similar analogies have been made for inflation \cite{Green:2014xqa,Amin:2015ftc} and gravity \cite{Rothstein:2012hk}.} All mass eigenstates are exponentially localized, and can thus have exponentially suppressed couplings to sites in the theory space --- the lightest (and heaviest) eigenstates especially so. In contrast to other mechanisms for generating exponential hierarchies in theory space, such as clockwork, all of the symmetries of the quiver are broken, so that this localization mechanism does not guarantee an exactly massless zero mode, but can guarantee light localized modes if desired.\footnote{In the case of the model of the following section, a sufficiently large number of sites can generate an arbitrarily light mode with arbitrarily suppressed couplings, with arbitrarily high confidence.} Also, unlike other theory space constructions, the coupling suppression is achieved without elaborate order across the quiver -- indeed, it is the very lack of order that leads to exponential hierarchies. This raises the compelling possibility that some of the large hierarchies observed in nature may result from disorder, rather than order. 

We demonstrate Anderson localization in theory space with a simple toy model for the localization of scalar fields, although the underlying mechanism is suitable for fields of any spin. We then construct two models in which the resulting exponential hierarchies explain small parameters of the Standard Model, namely neutrino masses and the quark flavor hierarchy. The mechanism is quite general and may be readily applied in a variety of contexts for physics beyond the Standard Model. 

\section{A toy model for the localization of scalar fields \label{sec:scalar}}

Consider a quadratic lagrangian of $N$ real scalars $\pi_i$ in 3+1 dimensions,
\begin{equation}
\mathcal{L}_\pi = \frac{1}{2} \sum_{i=1}^N (\partial \pi_i)^2 - \frac{1}{2} \sum_{i=1}^N \eps_i \pi_i^2 - \frac{1}{2} \sum_{i=1}^{N-1} t (\pi_i - \pi_{i+1})^2.
\label{eq:piLag}
\end{equation}
When viewed as a theory space lattice, the constant $t$ mediates mixing between adjacent sites, whereas the $\eps_i$ are site mass terms drawn randomly from a uniform distribution over the interval $[0,W]$.\footnote{Here the interval $[0,W]$ is chosen for simplicity to ensure that all masses-squared are positive, thereby avoiding questions of vacuum stability and facilitating generalization to higher spins. However, the localization effects of interest are insensitive to the choice of interval, and one could equally consider site mass terms drawn from a uniform distribution over $[-W/2,W/2]$ so that the average site mass is zero, in direct analogy with Anderson localization.} The combined mass matrix has $N$ positive eigenvalues $m_n^2$, whose corresponding eigenvectors $v^n_j$ are exponentially localized. To wit, away from the site $j_0$ where a given eigenvector has its greatest support, its $j$th components are well fit by the functional form $\abs{v^n_j} \propto \exp \left( \frac{-\abs{j-j_0}}{L_n} \right)$ for some localization length $L_n$. While the exponential localization of the eigenvectors is perhaps unsurprising for $W \gg t$, it persists even in the limit $W \ll t$. 

Moreover, the localization mechanism of (\ref{eq:piLag}) is robust both to turning on disorder in the hopping terms (even at the expense of turning it off in the site terms), and also to changing the distribution from which the random parameters are drawn. Any symmetric tridiagonal matrix $M_{i,j} = \eps_i \delta_{i,j} + t_i (\delta_{i +1,j} + \delta_{i,j+1})$, where the pairs $(\eps_i,t_i)$ are drawn independently from a two-dimensional distribution, will generically have localized eigenstates \cite{PhysRevB.4.396}, although there exist limiting cases where a measure-zero set of the eigenstates may be extended (see, e.g., \cite{PhysRevB.13.4597}). This is in contrast to when the values $\epsilon_i$ and $t_i$ are entirely deterministic. Absent some particular ordering of these parameters across the quiver, the eigenvectors will typically approximate the completely delocalized Bloch wave states obtained when there exists an exact cyclic symmetry, i.e., when $\exists k, \forall i,j, M_{i,j} = M_{i+k,j+k}$.

One such `particular ordering' is a clockwork lagrangian, which for $N$ scalars has quadratic part $\mathcal{L} = \frac{1}{2} \sum_{i=1}^N (\partial \pi_i)^2 - \frac{1}{2} m^2 \sum_{i=1}^{N-1} (\pi_i - q \pi_{i+1})^2$. Assuming $q>1$, there exists one localized massless state with localization length $\ln q$, and $N-1$ delocalized massive states. We compare this to the behavior of (\ref{eq:piLag}) below.

By working in the large $N$ limit, and considering an ensemble of the mass matrices (\ref{eq:piLag}), we may define an average localization length $L(m_i^2,t,W)$ as a function of the eigenvalue and the lagrangian parameters. Note that the mass matrix of (\ref{eq:piLag}) is purposefully almost identical to the Hamiltonian of the Anderson tight-binding model \cite{PhysRev.109.1492},
\begin{equation}
H_{i,j} = \eps_i \delta_{i,j} + t (\delta_{i+1,j} + \delta_{i,j+1}),
\label{eq:tightHam}
\end{equation}
with $\eps_i$ drawn from a uniform distribution over $[2t,W+2t]$.\footnote{The sole discrepancies lie in the $i=j=1$ and $i=j=N$ components, whose effects may be neglected in the large $N$ limit.} The mixing terms between sites correspond to hopping terms in the tight-binding model, while the random site mass terms correspond to local defect potentials. We recycle analytic expressions for $L$ from the tight-binding model literature \cite{IZRAILEV2012125}.

We identify two limiting behaviors of the lagrangian (\ref{eq:piLag}). One, the strong localization regime when $W \gg t$. If $t=0$, the eigenvalues are uniformly distributed over the interval $[0,W]$, and each eigenvector is perfectly localized. Perturbing about the $t=0$ state yields
\begin{equation}
L(m_i^2,t,W) \sim \left( \ln \frac{W}{2 t} - 1 \right)^{-1} .
\end{equation}

Two, the weak localization regime, when $W \ll t$. If $W=0$, the eigenstates are admixtures of delocalized plane waves, with a sinusoidal dispersion relation distributing eigenvalues over a mass-squared band $[0,4t]$. Turning on $W$, each plane wave is modulated by an exponential envelope, of size
\begin{equation}
L(m_i^2,t,W) \sim \begin{cases}
\left( \frac{t}{W} \right)^{\frac{2}{3}} & \text{if } \frac{2t - \abs{2t-m_n^2}}{W^\frac{4}{3} t^{-\frac{1}{3}}} \ll 1, \\
\frac{24}{W^2} (4 t^2 - (m_n^2 - 2 t)^2) & \text{otherwise.}\\
\end{cases}
\end{equation}
Note the anomalously strong localization of the states on the band edges.

\begin{figure}
\includegraphics[width=\columnwidth]{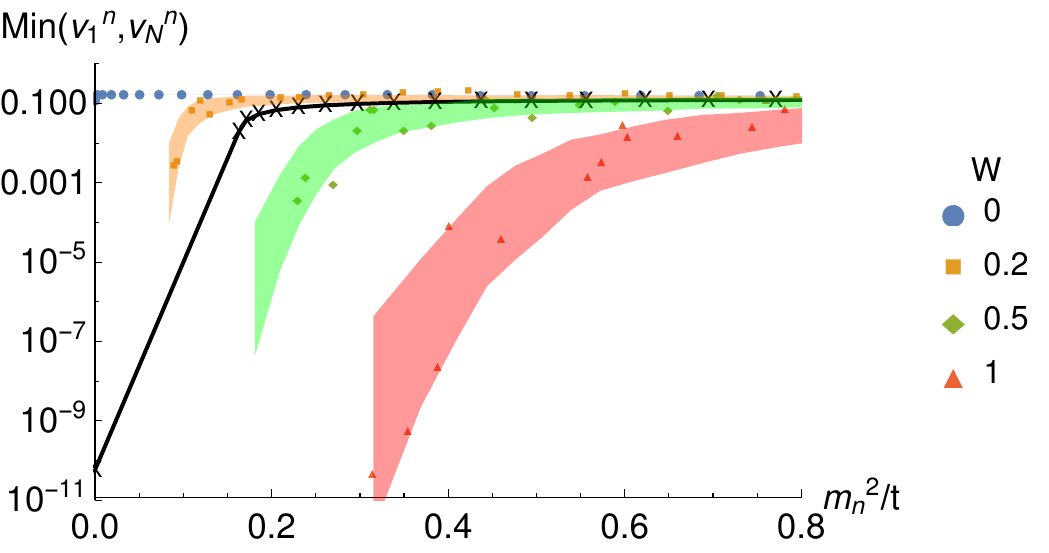}
\caption{The masses and smallest end-site overlaps of the light eigenmodes of (\ref{eq:piLag}) when $N=70$ and $t=1$. Shaded regions show typical values over an ensemble of matrices with given $W$, whereas points show masses and overlaps for a specific instance thereof. For comparison, black crosses show the masses and couplings of an $N=70$ `clockwork' lagrangian.\label{fig:andSpec}}
\end{figure}

Numerically we find that the above expressions for $L$ are $\cO(1)$ correct even for relatively small $N \approx 10$. To give an impression of the finite $N$ behavior of (\ref{eq:piLag}), for each eigenvector we plot the smallest of $v^n_1$ and $v^n_N$ against the corresponding mass $m_n^2$, shown in Fig.~\ref{fig:andSpec} for $N=70$, $t=1$ and a range of $W$ values. When $W=0$, there is no localization, and all states (including one massless state, and possibly several nearly massless states) have $\cO(\frac{1}{\sqrt{N}})$ overlaps with the end sites. Turning on $W$, the masses-squared shift by $\cO(W)$ and a mass gap appears. The end site overlaps of the lightest modes decrease: in the weak localization regime, the strongest effects are at the edge of the `band'. Further increasing $W$, all of the couplings descend markedly as the strong localization regime sets in. In black, Fig.~\ref{fig:andSpec} also shows the equivalent spectrum for a clockwork lagrangian of $q=1.4$ and $m^2 = t$: there exists one localized massless state with $\sim q^{-N}$ overlap with one end of the lattice, and $N-1$ largely delocalized states in a band beginning at mass $\sim m (q-1)$.

The above example may be cast in the more familiar language of a theory space with global symmetries associated to each site by beginning with $N$ complex scalars $\Phi_i$ and quadratic lagrangian
\begin{equation} \label{eq:phiLag}
\mathcal{L}_\Phi  = \sum_{i = 1}^N |\partial \Phi_i|^2 - \frac{1}{4} \sum_{i=1}^N \epsilon_i \Phi_i^2 - \sum_{i=1}^{N-1} t \Phi_i^\dag \Phi_{i+1} + {\rm h.c.} + \dots,
\end{equation}
where the $\dots$ include potential terms that preserve the underlying $U(1)^N$ global symmetry, while the terms proportional to $\epsilon$ and $t$ softly break $U(1)^N \rightarrow \varnothing$ and $U(1)^N \rightarrow U(1)$, respectively. Assuming the $U(1)^N$ preserving potential generates nonzero vacuum expectation values (vevs) $\langle \Phi_i \rangle$ at each site, then to quadratic order the lagrangian of the pseudo-goldstone bosons arising from (\ref{eq:phiLag}) is simply (\ref{eq:piLag}).

While we have illustrated the localization phenomenon for a toy model of scalar fields, analogous effects are possible for fermions and vector bosons. As we will demonstrate in the next section, for fermions both the hopping terms and the site masses may simply arise as terms in a Dirac mass matrix. For vector bosons, both hopping terms and site masses may arise from the vacuum expectation values of scalars transforming as bifundamentals and fundamentals, respectively. In each case, the localization properties of the eigenstates are akin to those of the scalar toy model.

\section{Anderson neutrino masses}

As a first application of Anderson localization in theory space to hierarchies in the Standard Model, we construct a model for realistic neutrino masses involving $\mathcal{O}(1)$ couplings and $\mathcal{O}$(TeV) scales in the underlying theory.   This is in contrast to conventional see-saw models in which small neutrino masses arise from a product of yukawas $y$ and Majorana neutrino masses $M$ satisfying $y^2 v / M \sim 10^{-11}$, implying a large input hierarchy of scales and/or couplings.

Consider a lattice of $N$ left-handed and $N$ right-handed Weyl fermions $L_i$ and $R_i$ with Dirac mass matrix (\ref{eq:tightHam}). Much like the `clockwork WIMP' model of \cite{Hambye:2016qkf}, a left-handed SM neutrino $\nu$ and right-handed Weyl fermion $\Psi$ mix through couplings at opposite ends of the lattice:
\begin{align}
\mathcal{L}_\text{mass} &= -t \bar{L_1} \Psi - \bar{L_i} H_{i,j} R_i - t \bar{\nu} R_N - W \Psi \Psi + \text{h.c.} \nonumber \\
&= -\sum_n \left( t v_1^n \bar{\psi^n_L} \Psi - m_n \bar{\psi^n_L} \psi^n_R - t v_N^n \bar{\nu} \psi^n_R \right) \nonumber \\
& \;\;\;\; - W \Psi \Psi + \text{h.c.} \, .
\end{align}
Via diagonalization of the matrix $H_{i,j}$, we have written the $L_i$ and $R_i$ as $N$ Dirac fermions $\psi^n$, masses $m_n$, each comprising a component $v^n_1$ of $L_1$ and a component $v^n_N$ of $R_N$. $\nu$ acquires a Majorana mass in a see-saw-like manner: diagrammatically, the $\nu$ mixes into a $\psi^n$, which mixes into $\Psi$ before returning to $\nu$ via another $\psi^m$. The neutrino mass
\begin{equation}
m_\nu \sim \left( \sum_{n=1}^N \frac{t v^n_N t v^n_1}{m_n} \right )^2 \frac{1}{W} \sim \left( \sum_{n=1}^N \frac{t^2}{m_n} e^{-\frac{N}{L_n}} \right )^2 \frac{1}{W},
\end{equation}
where $L_n$ is the localization length of the $n$th mode. If the $\eps_i$ are drawn uniformly from $[W,2W]$, and $t=\frac{2}{5}W$, then an $N=10$ lattice yields $m_\nu \sim 10^{0 \pm 1} \eV$ for $W = 1 \TeV$. A reasonably economical and generic set of \cO(TeV) parameters thereby reliably generate an \cO(eV) scale. 

The masses of the other fermion states of the lattice lie in the band $[0.5 \TeV,2 \TeV]$, some of which have $\cO(1)$ Yukawa couplings to the Higgs and left-handed leptons of the Standard Model, once electroweak symmetry is restored.  These states may thus give (lepton number conserving) signatures in sterile neutrino searches at collider experiments. In contrast to analogous models of neutrino masses in more orderly frameworks such as clockwork and warped extra dimensions, the couplings of these heavy states to the Standard Model can vary randomly from state to state within the expected envelope.

A similar setup may be used to construct an Anderson WIMP dark matter candidate with an exponentially long lifetime, in close analogy with the clockwork WIMP \cite{Hambye:2016qkf}.

\section{Anderson partial compositeness}

Anderson localization in theory space may also be fruitfully applied to Standard Model flavor hierarchies. Here we consider a purely four-dimensional model of partial compositeness \cite{Kaplan:1991dc} in which inter-generational hierarchies arise from the exponential profile of a localized scalar.

In partial compositeness models, each Standard Model (SM) quark (which we write in lower case, $\{q^i,u^i,d^i\}$, with generational indices $i=1,2,3$) has a vector-like partner (upper case $\{Q^i,U^i,D^i\}$) with the same gauge quantum numbers. A schematic partial compositeness lagrangian looks like
\begin{align}
\mathcal{L} \supset - M_\text{f} (\eta^i_q \bar{q}^i Q^i + \eta^i_u \bar{u}^i U^i + \eta^i_d \bar{d}^i D^i \nonumber \\
+ \bar{Q}^i Q^i + \bar{U}^i U^i + \bar{D}^i D^i ) \nonumber \\
- Y_u^{ij} \bar{Q}^i \tilde H U^j - Y_d^{ij} \bar{Q}^i H D^j + \text{h.c.},
\end{align}
where repeated generational indices are summed over. The entries of the Yukawa matrices $Y_u$ and $Y_d$ have anarchic values of order some overall constant $Y$, and $M_\text{f}$ is \cO(TeV). We arrange a hierarchy of $\eta$s such that the following parametric relations are satisfied for the observable masses and mixings of the SM quark sector \cite{KerenZur:2012fr}
\begin{equation}
m_u^i \sim \eta_q^i \eta_u^i Y v; \quad m_d^i \sim \eta_q^i \eta_d^i Y v; \quad V_\text{CKM}^{ij} \sim \frac{\eta_q^i}{\eta_q^j} \text{ if } i<j,
\label{eq:PCcons}
\end{equation}
$v$ being the Higgs vev.

Such hierarchies are typically explained by dynamics in the composite sector containing the vector-like partners. However, we can instead reproduce the desired pattern of $\eta$s purely from Anderson localization, using (for example) the vev of a quiver of $N=9$ SM singlet scalars $\pi_i$.  Consider the lagrangian of (\ref{eq:piLag}) with the addition of site quartic terms
\begin{equation}
\mathcal{L} = \mathcal{L}_\pi + \lambda \sum_{i=1}^N \pi_i^4.
\end{equation}
If the quadratic terms of $\mathcal{L}_\pi$ have negative eigenvalues, the $\pi$s will acquire a vev which lies mostly in the direction of the corresponding localized eigenvectors. The localization of the vev is particularly pronounced if there is exactly one negative mass eigenvalue.

For example, let $\lambda=1$, $t=0.4 M_\text{f}^2$ and draw $\eps_i$ uniformly from $[-0.6 M_\text{f}^2,1 M_\text{f}^2]$, such that about half the time there is exactly one negative eigenvalue. Couple each SM quark and its partner to a different $\pi_i$ via a Yukawa interaction, such that
\begin{equation}
 (\langle \pi_1 \rangle,\ldots,\langle \pi_9 \rangle) = M_\text{f} (\eta_u^A,\eta_d^A,\eta_q^A,\eta_u^B,\eta_d^B,\eta_d^C,\eta_q^B,\eta_u^C,\eta_q^C).
\end{equation}
Here $\{A,B,C\}$ is a particular permutation of the generational indices $\{1,2,3\}$, determined \emph{a posteriori} from the sizes of the $\eta_q^i$. About 5\% of the time, the two relations $0.1< \frac{\eta_q^1}{\eta_q^2}<0.3$ and $0.1^2< \frac{\eta_q^2}{\eta_q^3}<0.3^2$ are simultaneously satisfied, approximately reproducing the parametric hierarchy of the CKM matrix. These points also typically reproduce quark masses correct to within an order of magnitude for $Y \sim \cO(1)$. We thus obtain a viable model of partial compositeness in which the necessary exponential hierarchies are generated by Anderson localization in theory space, rather than dynamics of a strongly interacting composite sector.

In addition to vectorlike fermion states of mass $\sim M_\text{f}$, there exist physical excitations of the scalar lattice of mass $\cO(M_\text{f})$, some of which have \cO(1) flavor-changing Yukawa couplings to light quarks. The possibility of virtual tree-level exchange of these scalar particles in kaon mixing sets a rough phenomenological bound of $M_\text{f} \gtrsim 300 \TeV$. 

%The distributions of the products $\eps_q^i \eps_u^i$ --- proportional to the up quark masses --- are shown in Fig.\ref{fig:upMasses}. It's similar for down quark masses. $\frac{\eps_q^1}{\eps_q^2}$, which should be $\sim \sin \theta_c \sim 0.2$, is plotted in Fig.\ref{fig:wolfenstein}.

\section{Discussion \& Conclusions}

In this work we have identified a new mechanism for generating exponential hierarchies in four-dimensional quantum field theories, via an analogue of Anderson localization in theory space. In contrast to clockwork or deconstructions of extra-dimensional theories, exponential localization via disorder in theory space requires neither an elaborate ordering of parameters nor exponential hierarchies as inputs. Novel features include the exponential localization of all mass eigenstates; the anomalous localization of the lightest and heaviest mass eigenstates; and the statistical nature of mass eigenstate couplings to specific sites. We have further illustrated the potential for Anderson localization in theory space to explain observed hierarchies in nature by constructing simple models for neutrino and quark masses.

There are numerous directions for further inquiry. In this work we have focused on local theory spaces (i.e.~nearest-neighbor interactions) with diagonal disorder, but it would be fruitful to explore Anderson localization in more general theory spaces insofar as localization persists in both local theory spaces with off-diagonal disorder and classes of non-local theory spaces. Insofar as we have only sketched applications of the localization mechanism to neutrino masses and quark flavor textures, further study of their phenomenology and experimental signatures is warranted. Given that the localization mechanism can be applied to particles of any spin, it should also be possible to develop applications to a variety of other hierarchies in and beyond the Standard Model. As long as exactly massless zero modes are not required, it should be amenable to many of the same applications as generalized clockwork \cite{Giudice:2016yja}. 

In particular, it is natural to speculate whether this mechanism might be useful for addressing the electroweak hierarchy problem. The most obvious application is to a theory space of spin-two particles \cite{ArkaniHamed:2002sp}, thereby realizing the four-dimensional deconstruction of gravitational Anderson localization proposed in \cite{Rothstein:2012hk}. The exponential localization of the graviton zero mode would then translate to a warp factor controlling the scale of strong coupling across the theory space. However, the naive realization of Anderson localization in a spin-two theory space does not manifestly preserve a massless graviton or an unbroken diffeomorphism invariance. Given the significance of the hierarchy problem and the novelty of the localization mechanism, further study is clearly warranted. 

More broadly, both clockwork and the analogues of Anderson localization presented in this work underscore the relevance of lower-dimensional physics to model-building in higher-dimensional field theories. It would be worthwhile to pursue further analogies between diverse condensed matter systems and four-dimensional theory spaces. \\

\begin{acknowledgments}
We are grateful to Daniel Green and Scott Thomas for numerous useful conversations, and to Daniel Green for comments on the manuscript. This work is supported in part by the US Department of Energy under the grant DE- SC0014129.
\end{acknowledgments}

\bibliography{andersonbib}

\begin{thebibliography}{27}
\expandafter\ifx\csname natexlab\endcsname\relax\def\natexlab#1{#1}\fi
\expandafter\ifx\csname bibnamefont\endcsname\relax
  \def\bibnamefont#1{#1}\fi
\expandafter\ifx\csname bibfnamefont\endcsname\relax
  \def\bibfnamefont#1{#1}\fi
\expandafter\ifx\csname citenamefont\endcsname\relax
  \def\citenamefont#1{#1}\fi
\expandafter\ifx\csname url\endcsname\relax
  \def\url#1{\texttt{#1}}\fi
\expandafter\ifx\csname urlprefix\endcsname\relax\def\urlprefix{URL }\fi
\providecommand{\bibinfo}[2]{#2}
\providecommand{\eprint}[2][]{\url{#2}}

\bibitem[{\citenamefont{Dirac}(1937)}]{Dirac:1937ti}
\bibinfo{author}{\bibfnamefont{P.~A.~M.} \bibnamefont{Dirac}},
  \bibinfo{journal}{Nature} \textbf{\bibinfo{volume}{139}},
  \bibinfo{pages}{323} (\bibinfo{year}{1937}).

\bibitem[{\citenamefont{Dirac}(1938)}]{Dirac:1938mt}
\bibinfo{author}{\bibfnamefont{P.~A.~M.} \bibnamefont{Dirac}},
  \bibinfo{journal}{Proc. Roy. Soc. Lond.} \textbf{\bibinfo{volume}{A165}},
  \bibinfo{pages}{199} (\bibinfo{year}{1938}).

\bibitem[{\citenamefont{Weinberg}(1976)}]{Weinberg:1975gm}
\bibinfo{author}{\bibfnamefont{S.}~\bibnamefont{Weinberg}},
  \bibinfo{journal}{Phys. Rev.} \textbf{\bibinfo{volume}{D13}},
  \bibinfo{pages}{974} (\bibinfo{year}{1976}), \bibinfo{note}{[Addendum: Phys.
  Rev.D19,1277(1979)]}.

\bibitem[{\citenamefont{Susskind}(1979)}]{Susskind:1978ms}
\bibinfo{author}{\bibfnamefont{L.}~\bibnamefont{Susskind}},
  \bibinfo{journal}{Phys. Rev.} \textbf{\bibinfo{volume}{D20}},
  \bibinfo{pages}{2619} (\bibinfo{year}{1979}).

\bibitem[{\citenamefont{Grossman and Neubert}(2000)}]{Grossman:1999ra}
\bibinfo{author}{\bibfnamefont{Y.}~\bibnamefont{Grossman}} \bibnamefont{and}
  \bibinfo{author}{\bibfnamefont{M.}~\bibnamefont{Neubert}},
  \bibinfo{journal}{Phys. Lett.} \textbf{\bibinfo{volume}{B474}},
  \bibinfo{pages}{361} (\bibinfo{year}{2000}), \eprint{hep-ph/9912408}.

\bibitem[{\citenamefont{Gherghetta and Pomarol}(2000)}]{Gherghetta:2000qt}
\bibinfo{author}{\bibfnamefont{T.}~\bibnamefont{Gherghetta}} \bibnamefont{and}
  \bibinfo{author}{\bibfnamefont{A.}~\bibnamefont{Pomarol}},
  \bibinfo{journal}{Nucl. Phys.} \textbf{\bibinfo{volume}{B586}},
  \bibinfo{pages}{141} (\bibinfo{year}{2000}), \eprint{hep-ph/0003129}.

\bibitem[{\citenamefont{Arkani-Hamed
  et~al.}(2001{\natexlab{a}})\citenamefont{Arkani-Hamed, Cohen, and
  Georgi}}]{ArkaniHamed:2001nc}
\bibinfo{author}{\bibfnamefont{N.}~\bibnamefont{Arkani-Hamed}},
  \bibinfo{author}{\bibfnamefont{A.~G.} \bibnamefont{Cohen}}, \bibnamefont{and}
  \bibinfo{author}{\bibfnamefont{H.}~\bibnamefont{Georgi}},
  \bibinfo{journal}{Phys. Lett.} \textbf{\bibinfo{volume}{B513}},
  \bibinfo{pages}{232} (\bibinfo{year}{2001}{\natexlab{a}}),
  \eprint{hep-ph/0105239}.

\bibitem[{\citenamefont{Douglas and Moore}(1996)}]{Douglas:1996sw}
\bibinfo{author}{\bibfnamefont{M.~R.} \bibnamefont{Douglas}} \bibnamefont{and}
  \bibinfo{author}{\bibfnamefont{G.~W.} \bibnamefont{Moore}}
  (\bibinfo{year}{1996}), \eprint{hep-th/9603167}.

\bibitem[{\citenamefont{Georgi}(1986)}]{Georgi:1985hf}
\bibinfo{author}{\bibfnamefont{H.}~\bibnamefont{Georgi}},
  \bibinfo{journal}{Nucl. Phys.} \textbf{\bibinfo{volume}{B266}},
  \bibinfo{pages}{274} (\bibinfo{year}{1986}).

\bibitem[{\citenamefont{Arkani-Hamed
  et~al.}(2001{\natexlab{b}})\citenamefont{Arkani-Hamed, Cohen, and
  Georgi}}]{ArkaniHamed:2001ca}
\bibinfo{author}{\bibfnamefont{N.}~\bibnamefont{Arkani-Hamed}},
  \bibinfo{author}{\bibfnamefont{A.~G.} \bibnamefont{Cohen}}, \bibnamefont{and}
  \bibinfo{author}{\bibfnamefont{H.}~\bibnamefont{Georgi}},
  \bibinfo{journal}{Phys. Rev. Lett.} \textbf{\bibinfo{volume}{86}},
  \bibinfo{pages}{4757} (\bibinfo{year}{2001}{\natexlab{b}}),
  \eprint{hep-th/0104005}.

\bibitem[{\citenamefont{Hill et~al.}(2001)\citenamefont{Hill, Pokorski, and
  Wang}}]{Hill:2000mu}
\bibinfo{author}{\bibfnamefont{C.~T.} \bibnamefont{Hill}},
  \bibinfo{author}{\bibfnamefont{S.}~\bibnamefont{Pokorski}}, \bibnamefont{and}
  \bibinfo{author}{\bibfnamefont{J.}~\bibnamefont{Wang}},
  \bibinfo{journal}{Phys. Rev.} \textbf{\bibinfo{volume}{D64}},
  \bibinfo{pages}{105005} (\bibinfo{year}{2001}), \eprint{hep-th/0104035}.

\bibitem[{\citenamefont{Randall et~al.}(2003)\citenamefont{Randall, Shadmi, and
  Weiner}}]{Randall:2002qr}
\bibinfo{author}{\bibfnamefont{L.}~\bibnamefont{Randall}},
  \bibinfo{author}{\bibfnamefont{Y.}~\bibnamefont{Shadmi}}, \bibnamefont{and}
  \bibinfo{author}{\bibfnamefont{N.}~\bibnamefont{Weiner}},
  \bibinfo{journal}{JHEP} \textbf{\bibinfo{volume}{01}}, \bibinfo{pages}{055}
  (\bibinfo{year}{2003}), \eprint{hep-th/0208120}.

\bibitem[{\citenamefont{Giudice and McCullough}(2017)}]{Giudice:2016yja}
\bibinfo{author}{\bibfnamefont{G.~F.} \bibnamefont{Giudice}} \bibnamefont{and}
  \bibinfo{author}{\bibfnamefont{M.}~\bibnamefont{McCullough}},
  \bibinfo{journal}{JHEP} \textbf{\bibinfo{volume}{02}}, \bibinfo{pages}{036}
  (\bibinfo{year}{2017}), \eprint{1610.07962}.

\bibitem[{\citenamefont{Choi et~al.}(2014)\citenamefont{Choi, Kim, and
  Yun}}]{Choi:2014rja}
\bibinfo{author}{\bibfnamefont{K.}~\bibnamefont{Choi}},
  \bibinfo{author}{\bibfnamefont{H.}~\bibnamefont{Kim}}, \bibnamefont{and}
  \bibinfo{author}{\bibfnamefont{S.}~\bibnamefont{Yun}},
  \bibinfo{journal}{Phys. Rev.} \textbf{\bibinfo{volume}{D90}},
  \bibinfo{pages}{023545} (\bibinfo{year}{2014}), \eprint{1404.6209}.

\bibitem[{\citenamefont{Choi and Im}(2016)}]{Choi:2015fiu}
\bibinfo{author}{\bibfnamefont{K.}~\bibnamefont{Choi}} \bibnamefont{and}
  \bibinfo{author}{\bibfnamefont{S.~H.} \bibnamefont{Im}},
  \bibinfo{journal}{JHEP} \textbf{\bibinfo{volume}{01}}, \bibinfo{pages}{149}
  (\bibinfo{year}{2016}), \eprint{1511.00132}.

\bibitem[{\citenamefont{Kaplan and Rattazzi}(2016)}]{Kaplan:2015fuy}
\bibinfo{author}{\bibfnamefont{D.~E.} \bibnamefont{Kaplan}} \bibnamefont{and}
  \bibinfo{author}{\bibfnamefont{R.}~\bibnamefont{Rattazzi}},
  \bibinfo{journal}{Phys. Rev.} \textbf{\bibinfo{volume}{D93}},
  \bibinfo{pages}{085007} (\bibinfo{year}{2016}), \eprint{1511.01827}.

\bibitem[{\citenamefont{Anderson}(1958)}]{PhysRev.109.1492}
\bibinfo{author}{\bibfnamefont{P.~W.} \bibnamefont{Anderson}},
  \bibinfo{journal}{Phys. Rev.} \textbf{\bibinfo{volume}{109}},
  \bibinfo{pages}{1492} (\bibinfo{year}{1958}).

\bibitem[{\citenamefont{Green}(2015)}]{Green:2014xqa}
\bibinfo{author}{\bibfnamefont{D.}~\bibnamefont{Green}},
  \bibinfo{journal}{JCAP} \textbf{\bibinfo{volume}{1503}}, \bibinfo{pages}{020}
  (\bibinfo{year}{2015}), \eprint{1409.6698}.

\bibitem[{\citenamefont{Amin and Baumann}(2016)}]{Amin:2015ftc}
\bibinfo{author}{\bibfnamefont{M.~A.} \bibnamefont{Amin}} \bibnamefont{and}
  \bibinfo{author}{\bibfnamefont{D.}~\bibnamefont{Baumann}},
  \bibinfo{journal}{JCAP} \textbf{\bibinfo{volume}{1602}}, \bibinfo{pages}{045}
  (\bibinfo{year}{2016}), \eprint{1512.02637}.

\bibitem[{\citenamefont{Rothstein}(2013)}]{Rothstein:2012hk}
\bibinfo{author}{\bibfnamefont{I.~Z.} \bibnamefont{Rothstein}},
  \bibinfo{journal}{Phys. Rev. Lett.} \textbf{\bibinfo{volume}{110}},
  \bibinfo{pages}{011601} (\bibinfo{year}{2013}), \eprint{1211.7149}.

\bibitem[{\citenamefont{Economou and Cohen}(1971)}]{PhysRevB.4.396}
\bibinfo{author}{\bibfnamefont{E.~N.} \bibnamefont{Economou}} \bibnamefont{and}
  \bibinfo{author}{\bibfnamefont{M.~H.} \bibnamefont{Cohen}},
  \bibinfo{journal}{Phys. Rev. B} \textbf{\bibinfo{volume}{4}},
  \bibinfo{pages}{396} (\bibinfo{year}{1971}).

\bibitem[{\citenamefont{Theodorou and Cohen}(1976)}]{PhysRevB.13.4597}
\bibinfo{author}{\bibfnamefont{G.}~\bibnamefont{Theodorou}} \bibnamefont{and}
  \bibinfo{author}{\bibfnamefont{M.~H.} \bibnamefont{Cohen}},
  \bibinfo{journal}{Phys. Rev. B} \textbf{\bibinfo{volume}{13}},
  \bibinfo{pages}{4597} (\bibinfo{year}{1976}).

\bibitem[{\citenamefont{Izrailev et~al.}(2012)\citenamefont{Izrailev, Krokhin,
  and Makarov}}]{IZRAILEV2012125}
\bibinfo{author}{\bibfnamefont{F.}~\bibnamefont{Izrailev}},
  \bibinfo{author}{\bibfnamefont{A.}~\bibnamefont{Krokhin}}, \bibnamefont{and}
  \bibinfo{author}{\bibfnamefont{N.}~\bibnamefont{Makarov}},
  \bibinfo{journal}{Physics Reports} \textbf{\bibinfo{volume}{512}},
  \bibinfo{pages}{125 } (\bibinfo{year}{2012}).

\bibitem[{\citenamefont{Hambye et~al.}(2017)\citenamefont{Hambye, Teresi, and
  Tytgat}}]{Hambye:2016qkf}
\bibinfo{author}{\bibfnamefont{T.}~\bibnamefont{Hambye}},
  \bibinfo{author}{\bibfnamefont{D.}~\bibnamefont{Teresi}}, \bibnamefont{and}
  \bibinfo{author}{\bibfnamefont{M.~H.~G.} \bibnamefont{Tytgat}},
  \bibinfo{journal}{JHEP} \textbf{\bibinfo{volume}{07}}, \bibinfo{pages}{047}
  (\bibinfo{year}{2017}), \eprint{1612.06411}.

\bibitem[{\citenamefont{Kaplan}(1991)}]{Kaplan:1991dc}
\bibinfo{author}{\bibfnamefont{D.~B.} \bibnamefont{Kaplan}},
  \bibinfo{journal}{Nucl. Phys.} \textbf{\bibinfo{volume}{B365}},
  \bibinfo{pages}{259} (\bibinfo{year}{1991}).

\bibitem[{\citenamefont{Keren-Zur et~al.}(2013)\citenamefont{Keren-Zur, Lodone,
  Nardecchia, Pappadopulo, Rattazzi, and Vecchi}}]{KerenZur:2012fr}
\bibinfo{author}{\bibfnamefont{B.}~\bibnamefont{Keren-Zur}},
  \bibinfo{author}{\bibfnamefont{P.}~\bibnamefont{Lodone}},
  \bibinfo{author}{\bibfnamefont{M.}~\bibnamefont{Nardecchia}},
  \bibinfo{author}{\bibfnamefont{D.}~\bibnamefont{Pappadopulo}},
  \bibinfo{author}{\bibfnamefont{R.}~\bibnamefont{Rattazzi}}, \bibnamefont{and}
  \bibinfo{author}{\bibfnamefont{L.}~\bibnamefont{Vecchi}},
  \bibinfo{journal}{Nucl. Phys.} \textbf{\bibinfo{volume}{B867}},
  \bibinfo{pages}{394} (\bibinfo{year}{2013}), \eprint{1205.5803}.

\bibitem[{\citenamefont{Arkani-Hamed et~al.}(2003)\citenamefont{Arkani-Hamed,
  Georgi, and Schwartz}}]{ArkaniHamed:2002sp}
\bibinfo{author}{\bibfnamefont{N.}~\bibnamefont{Arkani-Hamed}},
  \bibinfo{author}{\bibfnamefont{H.}~\bibnamefont{Georgi}}, \bibnamefont{and}
  \bibinfo{author}{\bibfnamefont{M.~D.} \bibnamefont{Schwartz}},
  \bibinfo{journal}{Annals Phys.} \textbf{\bibinfo{volume}{305}},
  \bibinfo{pages}{96} (\bibinfo{year}{2003}), \eprint{hep-th/0210184}.

\end{thebibliography}

\end{document}